\documentclass[aps,twocolumn,groupedaddress,floatfix,showpacs]{revtex4}

\usepackage{epsfig}
\usepackage{amssymb}
\usepackage{amsmath}

\begin{document}

\title{Quantum Smoluchowski equation: A systematic study}

\author{Stefan A. Maier\footnote{Permanent address: Institute for Theoretical Physics C, RWTH Aachen University, 52062 Aachen, Germany} %
 and Joachim Ankerhold}
\affiliation{Institut f\"ur Theoretische Physik, Universit\"at Ulm, Albert-Einstein-Allee 11, 89069 Ulm, Germany}

\date{\today}

\begin{abstract}
The strong friction regime at low temperatures is analyzed systematically starting from the formally exact path integral expression for the reduced dynamics. This quantum Smoluchowski regime allows for a type of semiclassical treatment in the inverse friction strength so that higher order quantum corrections to the original quantum Smoluchowski equation [PRL {\bf 87}, 086802 (2001), PRL {\bf 101}, 11903 (2008)] can be derived. Drift and diffusion coefficients are  determined by the equilibrium distribution in position and are directly related to the corresponding action of extremal paths and fluctuations around them. It is shown that the inclusion of higher order corrections reproduces the quantum enhancement above crossover for the decay rate out of a metastable well exactly.
\end{abstract}


\maketitle

\section{Introduction}
Real systems interact with surrounding reservoirs which typically contain a macroscopic number of degrees of freedom and thus constitute heat baths. As a consequence, phenomena such as energy relaxation, fluctuations, and dephasing influence the system dynamics. In the classical domain the corresponding theoretical description is well developed and based on Langevin equations or, equivalently, Fokker-Planck equations for  phase space distributions \cite{risken}. The issue of dissipative quantum system has been initiated in the 1960s in the context of nuclear magnetic resonance \cite{wangsness_1953,redfield_1957} and quantum optics \cite{gardiner} and since the 1980s has attracted substantial research in condensed matter physics \cite{weiss,breuer}. Most recently, it has regained considerable attention for quantum information processing, where noise appears as an undesirable, but yet unavoidable effect \cite{averin}. In these systems the coupling between the relevant system and environmental degrees of freedom is designed to be as weak as possible. In contrast, finite dissipation can also be constructive and lead to qualitatively new processes such as stochastic resonance \cite{gammaitoni_1998}, charge transfer in molecular structures \cite{garg_1985,jortner_1999}, or ratchet induced transport in biomolecules \cite{reimann_2002}.
The regime of strong friction is classically known as the Smoluchowski limit \cite{smoluchowski_1915} and corresponds to a separation of time scales between equilibration of momentum, which is fast, and equilibration of position, which is slow. The generalization to the low temperature quantum domain, only a few years ago given in \cite{ankerhold_2001}, shows that quantum fluctuations may appear at relatively elevated temperatures and substantially influence the dynamics. Since then this issue has triggered various applications for strongly condensed phase systems, see e.g.\ \cite{machura_2004,ankerhold_2004,ankerhold_2004b,pollak_2004,machura_2006,dajka_2007}.

Theoretically, the description of quantum Brownian motion possesses an exact solution within the path integral representation for the reduced density matrix \cite{weiss,grabert_qbm,breuer}. This expression reveals that due to the non-Markovian nature of quantum mechanical fluctuations a simple equation of motion for the reduced density does in general not exist. Progress can be made in the weak coupling regime addressed above, where powerful master equations have been derived and successfully applied e.g.\ in quantum optical systems \cite{gardiner}. With a typical system frequency denoted by $\omega_0$, the condition for this reduction reads $\hbar\gamma\ll \hbar\omega_0, k_{\rm B} T$, where $\gamma$ is a typical coupling strength between system and bath and $T$ is the temperature. In the opposite range $\hbar\gamma\gg \hbar\omega_0, k_{\rm B} T$ friction dominates such that, roughly speaking, the typical linewidth induced by the environment exceeds the bare line separation as well as the thermal energy. In this deep quantum domain, named quantum Smoluchowski range (QSR), the reduced dynamics is nearly classical, but with a substantial impact of quantum fluctuations. Since friction dominates, any approximate treatment must start from a formulation where the system-bath interaction is described non-perturbatively as  e.g. in the path integral representation. It turns out that within this formulation a semiclassical type of approximation applies and the reduced dynamics can equivalently be cast into an equation of motion for the marginal distribution in position, the so-called quantum Smoluchowski equation (QSE) \cite{pag_2000}. In leading order the quantum fluctuations in the QSE have been derived in \cite{ankerhold_2001}.

What has not been done yet, is a systematic analysis of higher order corrections to the original QSE. In principle, this is a formidable task as it requires a systematic expansion of the full path integral expression for the real-time dynamics. However, as already pointed out in \cite{ankerhold_2001}, if a time evolution equation for the position distribution $P$ exists at all, it must have the form of a continuity equation, i.e.\ $ \dot{P}={\cal L}P$. For the distribution in thermal equilibrium $P_\beta$ one then has ${\cal L}P_\beta=0$ which can also be seen as an equation for ${\cal L}$ provided $P_\beta$ is known. Higher order quantum corrections in ${\cal L}$ can thus be determined from a systematic approximation to the exact path integral expression in imaginary time of the reduced equilibrium density matrix. Additional dynamical corrections in ${\cal L}$ need then be analyzed in the time window $1/\gamma\ll t\ll\gamma/\omega_0^2$ only. The corresponding extended QSE covers the dynamics of the position distribution for strong friction from high temperatures $\gamma\hbar\beta\ll 1$ to low temperatures $\gamma\hbar\beta\gg 1$.

An alternative approach, recently proposed for the high temperature range $\gamma\hbar\beta\ll 1$ by Coffey and co-workers \cite{coffey_2007,coffey_2008,coffey_2009}, follows  a similar strategy but is based on the thermal Wigner distribution for the {\em uncoupled} system. While this allows to obtain the universal leading quantum correction, we show here also, that it is {\em not} a consistent procedure to treat higher order quantum corrections. The latter ones carry information about the system-bath coupling, which is absent in the bare thermal distribution.

The paper is organized as follows: In Sec.~\ref{sec:general} a brief review of the path integral representation is  given and the relevant regimes in parameter space are discussed qualitatively. Then, in Sec.~\ref{sec:harmonic} the generic  case of systems with harmonic potential is studied for which the real-time dynamics can be solved exactly. This allows for a detailed analysis in the strong friction range at low temperatures. The extension to anharmonic systems is presented in Sec.~\ref{sec:anharmonic}, where the perturbation theory in the inverse friction strength is applied to derive higher order quantum corrections to the Moyal coefficients in the QSE. Interestingly, this type of semiclassical analysis also provides a formal solution to the problem which reveals the role of minimal action paths and deviations around them. As an application the escape rate out of a metastable well is calculated and shown to reproduce in higher order perturbation theory the exact result for the quantum enhancement factor above the crossover temperature. The paper concludes with a discussion of alternative approaches in Sec.~\ref{sec:comparison} and a summary of the results in Sec.~\ref{sec:discussion}.

\section{Preliminaries}\label{sec:general}

\subsection{Dissipative quantum dynamics}
\label{dissdyn}

Dissipative quantum systems are described with system+reservoir models \cite{caldeira_1983,weiss}, where the position $ q $ of a system with potential $ V (q) $
 is bilinearly coupled to the positions $ x_\alpha $ of environmental oscillators. Therefore, the Hamiltonian reads
\begin{align} \nonumber
 H & = H_S + H_B + H_I \\ \nonumber
 H_S & = \frac{p^2}{2m} + V (q) \\ \nonumber
 H_B & = \sum_\alpha \frac{p^2_\alpha}{2m_\alpha} + \frac{m_\alpha \omega^2_\alpha}{2} x_\alpha^2 \\
 H_I & =  \sum_\alpha \left[ - c_\alpha  q x_\alpha + \frac{c_\alpha^2}{m_\alpha \omega_\alpha^2} q^2 \right] \, ,
\end{align}
 with the $ q^2 $-dependent term in $ H_I $ added in order to avoid coupling-induced potential renormalizations. The dynamics of the full system described by a density operator $W(t)$ is then given by
\begin{equation}
 W (t) = {\rm e}^{- \frac{i}{\hbar} H t} W (0) {\rm e}^{+ \frac{i}{\hbar} H t}\, .
\end{equation}
The initial state $ W (0) $ is obtained from the equilibrium density operator $ W_\beta $
through the application of  projection operators acting on the Hilbert space of the system only.
In contrast to the Feynman-Vernon theory, it therefore bears initial correlations between the system and the bath. We are, however, only interested in the  reduced system  described by the density operator
 $ \rho (t) = {\rm tr}_{\rm B}\{W(t)\} $ and particularly focus on its position representation  $\langle q| \rho|q'\rangle = \rho (q,q') $. Further,  the analysis will be restricted to a class of initial preparations given by
\begin{equation}
 \rho(q_i,q_i',t=0) = \rho_\beta (q_i,q_i') \lambda(q_i,q_i')\, ,
\end{equation}
with a two-variable preparation function $ \lambda (q_i,q_i') $. Stemming from projection operators, it describes deviations from the reduced thermal equilibrium density $\rho_\beta={\rm tr}_B\{\exp(-\beta H)\}/(Z Z_B)$ with the bath partition function $Z_B$ and a proper normalization for the system $Z$.
 The path integral approach allows for an exact elimination of the bath degrees of freedom in the position representation.
 In doing so, one obtains
 \begin{align} \label{eq:dis_dyn}
  \rho (q_f,q_f',t) = & \int\int  dq_i dq_i'  J (q_f,q_f',t,q_i,q_i') \lambda (q_i,q'_i) \, .
 \end{align}
 The propagation function
  $ J (q_f,q_f',t,q_i,q_i') $ contains a threefold path integral over the system coordinates, two for the real-time propagation of the initial density matrix,
 one thermal path in imaginary time for the initial state.  For further details we refer to the literature \cite{grabert_qbm,weiss}.
 In time $ t $, the real time paths run from
 $ q_i $ and $ q_i' $ to $ q_f $ and $ q_f' $, respectively. On the imaginary time axis, the thermal path runs from $ q_i $ at time $ 0 $ to $ q_i' $ at time
 $ -i \hbar \beta $, where $ \beta = \left( k_B T \right)^{-1} $.
 Besides the bare propagation, the integrand in the propagation function contains an influence functional that keeps track of the interaction with the bath.
 The latter one is nonlocal in time and contains the damping kernel
\begin{equation}
 K( \theta ) = \int_0^\infty \frac{d\omega}{\pi} I (\omega) \frac{\cosh \left[ \omega (\hbar \beta /2 - i \theta )\right]}{\sinh ( \hbar \beta \omega /2)}  \, ,
\end{equation}
which is proportional to the force-force correlation function of the bath. It is determined by the spectral density
\begin{equation}
 I (\omega) = \pi \sum_\alpha \frac{c_\alpha^2}{2 m_\alpha \omega_\alpha} \delta \left( \omega - \omega_\alpha \right) .
\end{equation}
For purely real time arguments, this kernel reduces to
\begin{align} \label{eq:real_kernel}
 K (t) & = \int_0^\infty \frac{d\omega}{\pi} I (\omega) \left[ \coth (\hbar \beta \omega /2  ) \cos (\omega t) - i \sin (\omega t)  \right]\, ,
\end{align}
  and for imaginary times one finds  $K(-i\tau)=\mu~:~\delta(\tau)~:-k(\tau)$ with a time local contribution $\mu=(2/\pi)\int d\omega I(\omega)/\omega$. The time nonlocal one has a representation in terms of Matsubara frequencies $\nu_n=2\pi n/\hbar\beta$, i.e.,
\begin{equation}
\label{imkernel}
 k(\tau)  = \frac{m}{\hbar\beta} \sum_{n=-\infty}^\infty \hat{\gamma}(|\nu_n|)\, |\nu_n|\,{\rm e}^{i\nu_n\tau}\, ,
\end{equation}
where $\hat{\gamma}(z)$ denotes the Laplace transform of the classical friction kernel $\gamma(t)$.

 In the sequel, we consider Ohmic damping, i.e. a spectral density of the form $ I (\omega) =  m \tilde{\gamma} \omega $ and employ a Drude regularization, $ I (\omega) = m \tilde{\gamma} \omega \omega_c^2 / (\omega^2 + \omega_c^2) $ with a high frequency cut-off frequency $ \omega_c $, whenever needed.

\subsection{Regimes in  parameter space} \label{sec:par_reg}

While the path integral formulation provides with (\ref{eq:dis_dyn}) a formally exact expression for the reduced density matrix, its explicit evaluation is very demanding. Analytical results have been derived for harmonic systems, but in general even numerical approaches are limited to specific models or regimes in parameter space. It has further been shown that an equivalent ''simple'', i.e. tractable, equation of motion for the reduced density does not exist (see e.g.\ \cite{karrlein_1997}). The reason for these complications is the non-Markovian nature of quantum Brownian motion as determined by the time nonlocal damping kernel (\ref{eq:real_kernel}). In this situation progress can be made at least in a perturbative sense in certain ranges of parameter space. In this section we discuss the two complementary domains of weak and strong friction, respectively.

For this purpose,
a typical damping strength in the long time limit is introduced as
\begin{equation}
 \gamma \equiv \hat{\gamma} (0) = \lim_{\omega \rightarrow 0} \frac{I(\omega)}{m \omega} \, ,
\end{equation}
such that both for Ohmic and for Drude damping we have $ \gamma = \tilde{\gamma} $.
A second relevant frequency scale of the bath is temperature, i.e. $\nu_1=2\pi/\hbar\beta$, while for the isolated system we assume
 a typical frequency $ \omega_0 $. In fact, these three scales  define qualitatively the nature of the reduced dynamics.

In case of weak friction $\gamma / \omega_0 \ll 1 $, the typical relaxation time towards thermal equilibrium is of the order of $t_r\propto 1/\gamma$ and thus much larger than  any time scale of the bare system dynamics. As far as we are interested in phenomena like decoherence and dephasing, the damping kernel (\ref{eq:real_kernel}) becomes time-local on a coarse grained time scale $t\gg \hbar\beta$  provided that $\gamma\hbar\beta\ll 1$. This latter condition is the basic assumption for all types of master equations which have been used in fields like quantum optics, nuclear magnetic resonance, and most recently in quantum information processing.

In the opposite domain of strong friction $ \gamma/\omega_0 \gg 1 $ again a time scale separation exists, since the relaxation of position happens to occur for times on the order of $t_r\propto \gamma/\omega_0^2$, while the momentum relaxes on the time scale $1/\gamma$. Accordingly, a coarse graining of time leading to a time-local kernel (\ref{eq:real_kernel}) is possible if 
\begin{equation}
\label{condi}
 \hbar\beta, \frac{1}{\gamma},\frac{1}{\omega_c}\ll \frac{\gamma}{\omega_0^2}\, .
 \end{equation}
 On the one hand this latter relation comprises the classical regime $\gamma\hbar\beta\ll 1$ corresponding for strong friction to $\omega_0\hbar\beta\ll 1$, and on the other hand the deep quantum regime $\gamma\hbar\beta \gg 1$.

In the  classical domain of strong friction the Fokker-Planck equation for the distribution in full phase-space reduces to the famous Smoluchowski equation for the marginal distribution in position $P(q,t)=\langle q|\rho(t)|q\rangle$ \cite{risken,skinner_1979}, namely,
\begin{equation}
\label{class_smolu}
 \dot{P} (q,t) = \frac{1}{m \gamma} \partial_q \left[ V'(q) + \frac{1}{\beta} \partial_q \right] P(q,t) \, .
\end{equation}
 The quantum regime (Quantum Smoluchowski regime, QSR) has gained much attention only recently after a quantum Smoluchowski equation (QSE) has been derived from the path integral expression in \cite{ankerhold_2001}. The aim of this paper is to systematically derive higher order corrections to the original QSE in order to better understand its limitations, but also to obtain improved results.

\section{Harmonic oscillator}\label{sec:harmonic}

As mentioned above, harmonic systems $ V(q) = m \omega_0^2 q^2 /2 $ allow for an exact solution of the path integral expression (\ref{eq:dis_dyn}). They may thus serve as models to analyze the QSR and the existence of a QSE in detail.

\subsection{Time-dependent density matrix and time evolution equation} \label{sec:time_evo}

According to \cite{grabert_qbm} the propagating function for the reduced dynamics can be calculated explicitly and the reduced density matrix (\ref{eq:dis_dyn}) follows from
\begin{align}\nonumber
\label{finaldensity}
\rho(r_f, x_f, t)&=\frac{1}{4 \pi |A(t)|\sqrt{2\pi\langle q^2\rangle}} \int dx_i dr_i \lambda(r_i,x_i) \\
&\hspace{2cm}\times {\rm e}^{i\Sigma(r_f,x_f,t,r_i,x_i)/\hbar}\ ,
\end{align}
where we introduced sum and difference coordinates $ r= ( q+ q')/2 $ and $ x= q - q' $, respectively.
Here, the minimal action $\Sigma(\cdot)$ is of Gaussian form in the coordinates. It can be expressed in terms of the symmetric and antisymmetric part of the position autocorrelation function
\begin{align}
 S(t) = {\rm Re}\{\langle q(t) q \rangle_\beta\}\, , \quad A(t) = {\rm Im}\{\langle q(t) q \rangle_\beta\}\,
\end{align}
which are determined by the natural frequencies $\gamma/2\pm\sqrt{\gamma^2/4-\omega_0^2}$.
Further, the mean square of position in thermal equilibrium is found to read
\begin{equation}
\label{eq:harm_variances}
 \langle{q^2 }\rangle  =  \frac{1}{m \beta} \sum_{n=- \infty}^{+ \infty} \frac{1}{\nu_n^2 +  |\nu_n| \gamma + \omega_0^2}\ .
\end{equation}

In general, progress can now only be made upon specifying the preparation function in (\ref{finaldensity}) explicitly. However, simplifications arise in the strong friction domain considered here. Namely, for large friction $\gamma/\omega_0\gg 1$ and on the coarse grained time scale [cf.\ (\ref{condi})]
\begin{equation} \label{eq:smol_reg}
\hbar \beta, \frac{1}{\gamma}, \frac{1}{\omega_c} \ll t
\end{equation}
only the frequency
\begin{equation}
\label{eq:harm_parameter}
\Omega=\frac{\gamma}{2}-\sqrt{\frac{\gamma^2}{4}-\omega_0^2}
\end{equation}
is relevant.
In this parameter regime, the position autocorrelation functions reduce to
\begin{align}   \nonumber
 S(t) & = \frac{\hbar}{4m (\gamma/2-\Omega)} \cot \left(\hbar \beta \Omega / 2 \right) {\rm e}^{- \Omega t} \\
 A(t) & = - \frac{\hbar}{2m (\gamma/2-\Omega)}\, {\rm e}^{- \Omega t} \,
\end{align}
so that e.g.\ $A$ is of order $1/\gamma$ or smaller since $\gamma/2-\Omega\approx \gamma/2$. Further, in leading order the variance in position reduces to its classical value $\langle q^2\rangle\approx 1/m\beta\omega_0^2$.
Then, a straightforward analysis reveals that the Gaussian factor in $\Sigma$ restricts $x_i$ to be at most of order $A/\sqrt{\langle q^2\rangle}\sim 1/\sqrt{\gamma^2\langle q^2\rangle}$. Accordingly,
the $x_i$-dependence of the preparation function is only probed on this short length scale and we may set $\lambda(x_i,r_i)\approx \lambda(0,r_i)$ in (\ref{finaldensity}) provided the initial preparation is sufficiently smooth in $x_i$. Equivalently, the initial preparation must be restricted to momenta of order $\gamma\sqrt{\langle q^2\rangle}$ or smaller. This, however, is in complete accordance with the Smoluchowski limit which requires equilibration of momentum on sufficiently short time resp.\ length scales.

After performing now the $x_i$-integration and keeping the dominant terms only one arrives at
\begin{align}  \nonumber
 \rho(r_f,x_f,t) = &  \frac{{\rm e}^{- x_f^2 \langle{p^2}\rangle/2\hbar^2}}{\sqrt{2 \pi \Delta}}  \int d r_i\,  \lambda(r_i,0) P_\beta (r_i) \\ \label{eq:smol_dyn}
               & \times \exp\left\{- \frac{1}{2 \Delta}  \left[ r_f - r_i \frac{S (t) }{ \langle{q^2}\rangle} \right]^2\right\} ,
\end{align}
where
\begin{equation}
  \Delta = \langle q^2\rangle \left[ 1 - \frac{S^2 (t)}{\langle{q^2}\rangle^2} \right] \,
\end{equation}
and $P_\beta(r_i)=\rho_\beta(r_i,x_i=0)$ denotes the diagonal part of the reduced thermal distribution. The equilibrium variance in momentum is given by
\begin{equation}
\label{momentum}
\langle{p^2}\rangle  =  \frac{m}{ \beta} \sum_{n=- \infty}^{+ \infty} \frac{\gamma |\nu_n| + \omega_0^2}{\nu_n^2 +  |\nu_n| \gamma + \omega_0^2} \, .
\end{equation}
This sum  must be regularized by introducing a cut-off frequency $\omega_c$ as discussed above. In the interesting range $\gamma\hbar\beta\gg 1$ we then easily find $\langle p^2\rangle\approx (m\hbar\gamma/\pi){\rm log}(\omega_c/\gamma)$, while in the complementary high temperature regime $\gamma\hbar\beta\ll 1$ one regains the classical results   $\langle p^2\rangle\approx (m/\beta) [1+(\hbar\beta\gamma/\pi){\rm ln}(\hbar\beta\omega_c/2\pi)]$.
The leading corrections to the result (\ref{eq:smol_dyn}) consist of a phase factor $\exp[(i/\hbar) x_f \delta p (r_f,r_i)]$ with
\begin{equation}
\label{nonfaccorr}
 \delta p (r_f,r_i) =  \frac{\dot{A} (t)}{A (t)}\frac{m \langle q^2\rangle}{\Delta}\left[r_f - r_i \frac{S(t)}{\langle q^2\rangle}\right]\, .
 \end{equation}
For strong friction one has $\Omega\approx \omega_0^2/\gamma$ so that $\dot{A}/A\approx \omega_0^2/\gamma$. The order of magnitude of the factor in square brackets is given by the width of the Gaussian in (\ref{eq:smol_dyn}) as $\sqrt{\Delta}$ and that of $x_f$ by the Gaussian prefactor as $1/ \sqrt{\langle p^2\rangle}$. Hence, the above phase factor produces corrections in the exponential on the  order of $1/\sqrt{\gamma^2\langle p^2\rangle}$ or smaller. These corrections depend on the cut-off parameter $\omega_c$ and can be made arbitrarily small. Other corrections to (\ref{eq:smol_dyn}) are at most of order $\exp(-\gamma t)$ and are negligible on the coarse grained time scale.

Now,  a  Wigner transform of (\ref{eq:smol_dyn}) immediately gives rise to a momentum distribution proportional to $\exp(-p^2/2\langle p^2\rangle)$ as expected. For the relevant diagonal part $P(q,t)\equiv \rho(r_f,x_f=0,t)$ the dynamics can then be cast into an equation of motion for its time evolution \cite{pechukas_2001}
\begin{equation} \label{eq:ho_qse}
 \dot{P}  (q,t) = \frac{\Omega}{m\omega_0^2} \partial_q \left[D_1(q) + \partial_q D_2\right] P (q,t)
\end{equation}
 with a drift coefficient $D_1(q)=m\omega_0^2 q$ and a position independent diffusion term $D_2=m\omega_0^2\langle q^2\rangle$.
To leading order one regains  from the above expression in the QSR the classical Smoluchowski equation (\ref{class_smolu}) with $D_{2,\rm cl}=1/\beta$ and $\Omega/m\omega_0^2 \approx 1/m\gamma$. A systematic expansion around this result will be given in Sec.~\ref{sec:harm_perturb}.
Moreover, for systems driven externally by time dependent forces $f(q,t)$ the analysis goes through accordingly by replacing $m\omega_0^2 q\to m\omega_0^2 q+f(q,t)$ provided the typical time scale for the driving sufficiently exceeds the scales $1/\gamma, \hbar\beta$. This result has also been derived from a Quantum Langevin equation in \cite{dillenschneider_2009}.

\subsection{Initial correlations}\label{inicorr}

The pioneering work on quantum Brownian motion by Feynman and Vernon uses a factorized initial state, where the initial density of the whole system takes the form $W(0)=\rho_S(0) \exp(-\beta H_B)/Z_B$. Initial correlations between system
and reservoir are thus absent. Asymptotically, for very long times the reduced density matrix approaches the true reduced thermal equilibrium distribution as well, but for intermediate times factorizing and non-factorizing initial conditions lead in general to different stochastic processes. It is known that for high temperatures/weak friction both processes coincide after a transient period of time which is short compared to the relevant dynamics \cite{weiss,breuer}. Question here is if this holds true also for low temperatures/strong friction  so that the QSE (\ref{eq:ho_qse}) can also be derived within the somewhat simpler factorizing formulation.

Following the lines described in the previous section  one gains for factorized initial states a result which looks similar to (\ref{eq:smol_dyn}) with the replacements $\lambda(0,r_i)P_\beta(r_i)\to \rho_S(r_i,t=0)$,  $S(t)/\langle q^2\rangle\to -2m A(t)/\hbar$, and $\Delta\to \tilde{\Delta}=4 A^2\langle p^2\rangle/\hbar^2+\Delta$. In contrast to (\ref{eq:smol_dyn}), however, an additional phase factor $\exp[(i/\hbar)x_f r_f m \dot{A}/A]$ appears, which cannot be neglected.
Namely, while the order of magnitude of the corresponding phase factor  (\ref{nonfaccorr}) for non-factorized initial states is completely determined by the widths of the Gaussians in $x_f$ and $r_f-r_i S/\langle q^2\rangle$, respectively, this is not the case here. The order of magnitude of this term depends via $r_f$ also on the distribution of initial values for $r_i$. Hence, the density matrix does not reduce to a simple product of thermal momentum distribution and time dependent position distribution.

If one ignores this complication and focuses on the diagonal part of the density $\rho(q, x_f=0)$ only,  a time evolution equation of the form (\ref{eq:ho_qse}) is found, however, with a diffusion coefficient which directly depends on the momentum variance via $\langle p^2\rangle A(t)^2/\langle q^2\rangle$. While this term is negligible in the classical domain (it is then of order $1/\gamma^2$) and for times $t\gg \gamma/\omega_0^2$,  it provides an essential contribution for $\gamma\hbar\beta\gg 1$ on the coarse grained time scale (\ref{eq:smol_reg}). The result is a Smoluchowski type of equation  which contains via a time dependent drift coefficient  the momentum variance.
This analysis reveals that the formulation based on factorized initial states does  {\em not} lead to an acceptable  Smoluchowski description in the deep quantum domain. The reason for this behavior is that for strong friction the time scale on which initial correlations are established is identical to the relaxation time of the full system, i.e.\ equilibration of position.

\subsection{Equilibrium distribution and current operator}\label{harm_current}

 Since the treatment of the dynamics laid out in Sec.~\ref{sec:time_evo} requires ergodicity, a time evolution equation for the reduced density must describe the relaxation of the system
 to thermal equilibrium. As we will see here, this allows for an alternative derivation of quantum corrections to the Smoluchowski equation starting from the equilibrium density matrix of the open system.

 The time evolution equation (\ref{eq:ho_qse}) has the structure of a continuity equation for the marginal probability distribution in position.
 Accordingly, the thermal distribution $P_\beta$ of the harmonic system is determined by a vanishing current $ J = {\cal L} P_\beta (q)=0 $ with a current operator being defined as
\begin{equation}
\label{eq_dia_im}
 {\cal L} = D_1(q)  + \partial_q D_2 \,
\end{equation}
with Moyal coefficients $D_1$ and $D_2$. Note that only for systems with at most quadratic potentials $D_2$ is a constant.

The idea is now to consider (\ref{eq_dia_im}) as an {\em ansatz} with yet unknown Moyal coefficients. The thermal distribution has thus to be of the form
\begin{equation}
\label{eq:equ_sol}
 \hat{P}_\beta (q)  = \frac{1}{Z_0}\frac{{\rm e}^{- \psi(q)}}{D_2}\, , \ \psi (q)   = \int_0^q dy \frac{D_1 (y)}{D_2}
\end{equation}
with $Z_0$ being the partition function of the oscillator.

On the other hand, the reduced thermal density matrix for a general system
can be represented as a path integral in imaginary time, namely,
 \begin{equation}
 \rho_\beta(q,q')=\frac{1}{Z}\int {\cal D}[\bar{q}]\,  {\rm e}^{-S_{E, eff}[\bar{q}]/\hbar}\, ,
 \label{eq:impath}
 \end{equation}
 where $\bar{q}(\tau)$ connects in the time interval $\hbar\beta$ end-points $\bar{q}(0)=q'$ with $\bar{q}(\hbar\beta)=q$.
 The effective action contains the bare euclidian action of the system and the influence functional in imaginary time determined by the kernel (\ref{imkernel}). $Z$ denotes a proper normalization, which in case of a harmonic system coincides with the partition function.
 For quadratic potentials this expression can be evaluated exactly to read
\begin{equation}
\label{eq:density_im}
 \rho_\beta (r,x) = \frac{1}{\sqrt{2 \pi \langle{q^2}\rangle}} \exp\left[- \left( \frac{r^2}{2 \langle{q^2}\rangle} + \frac{x^2 \langle{p^2}\rangle}{2 \hbar^2} \right) \right]
\end{equation}
with the variances in position and momentum, respectively, given in (\ref{eq:harm_variances}) and (\ref{momentum}), respectively.
 Upon comparing $\hat{P}_\beta$ with the explicit result for $P_\beta(q)=\rho_\beta(q,0)$ one can determine $D_1$ and $D_2$. As expected, for the harmonic oscillator one regains the results specified already in (\ref{eq:ho_qse}).

This alternative procedure reveals an important feature. We can distinguish two types of corrections to the classical Smoluchowski equation: Those in the current operator are completely determined by the equilibrium properties of the system and are influenced by quantum fluctuations; the other ones are dynamical in origin and appear merely as an overall factor in front of $\partial_q {\cal L}$ [cf.~(\ref{eq:ho_qse})] completely determined by  classical dynamics.
Hence, as long as we are interested in the role of quantum fluctuations in the QSE for anharmonic systems, it may be justified  to perturbatively calculate the Moyal coefficients from the corresponding thermal distribution and assume that quantum corrections in the dynamics are relevant only in higher orders of perturbation theory.
In fact, the analysis presented first in \cite{ankerhold_2001} proves this scheme to be correct at least for the dominating quantum fluctuations. To prove that it applies also to higher order contributions requires to consider the reduced dynamics only within the time window $1/\gamma, \hbar\beta \ll t\ll \gamma/\omega_0^2$.

\subsection{Perturbation theory for strong friction}\label{sec:harm_perturb}

So far the QSE obtained in (\ref{eq:ho_qse}) is exact on the coarse grained time scale. To lay the basis for a perturbative treatment in case of anharmonic potentials, we study in the sequel a systematic expansion around this exact expression in terms of the small parameter $\omega_0/\gamma$. The corresponding results are thus valid for all temperatures provided damping is strong. For this purpose, one writes
\begin{align} \nonumber
 \langle{q^2}\rangle &= \langle{q^2}\rangle_{\rm cl} + \Lambda \\ \label{eq:exact_Lambda}
  \Lambda & = \frac{2}{m\beta} \sum_{n=1}^\infty \frac{1}{\nu_n^2 + \nu_n \gamma + \omega_0^2} \, ,
\end{align}
where $ \langle{q^2}\rangle_{\rm cl} = 1 /(\beta m \omega_0^2) $ is the classical variance in position.
For strong friction $\gamma/\omega_0\gg 1$ one then has
\begin{align}
 \Lambda & = \sum_{\mu=0}^\infty \frac{\omega_0^{2 \mu}}{\mu!} \lambda_\mu\nonumber \\
 \lambda_\mu & = \frac{2}{m \beta} \left. \partial_x^\mu \sum_{n=1}^\infty \left( \nu_n^2 + \gamma \nu_n + x \right)^{-1} \right|_{x=0} \, .
 \label{eq:Lambda}
\end{align}
 The coefficients $ \lambda_\mu $ are independent of system properties and can be expressed with the help of the polygamma functions. In lowest orders they read
\begin{align} \nonumber
  \lambda_0  = &  \frac{\hbar}{\pi m \gamma} \left[ \Psi \left( 1 + \frac{\gamma}{\nu} \right) + C_E \right]  \, ,\\ \nonumber
 \lambda_1  =  & \frac{2}{m \beta \nu^2 \gamma^2 }  \left\{ \frac{2 \nu}{\gamma} \left[ \Psi \left( 1+ \frac{\gamma}{\nu} \right) + C_E \right]  \right. \\
  &-  \left.  \Psi^{(1)} \left( 1+ \frac{\gamma}{\nu} \right) - \frac{\pi^2}{6}  \right\}\,  ,
\end{align}
with $ C_E$ being the Euler-Mascheroni constant and $\Psi^{(k)}$ the $k$th derivative of the digamma function $\Psi$.  In particular, one has in the high temperature domain $\gamma\hbar\beta\ll 1$ the approximation $\lambda_0\approx  \hbar^2 \beta / 12 m$, while in the quantum regime $\gamma\hbar\beta\gg 1$ the expression $\lambda_0\approx (\hbar/\pi m\gamma) {\rm ln}(\gamma\hbar\beta/2\pi)$ reveals that quantum fluctuations are substantial and not algebraically small. The coefficients $ \lambda_\mu, \mu\geq 1 $ are on order of ${\rm ln} (\gamma)/\gamma^{2\mu+1}$ and thus serve as semiclassical expansion parameters. We note that their dependence on Planck's constant is highly non-algebraic.
 This way,  one obtains the leading quantum corrections in (\ref{eq:ho_qse}) as
 \begin{equation} \label{eq:ho_qse_pertur}
 \dot{P}  (q,t) = \frac{1}{m\gamma} \partial_q \left[m\omega_0^2 q + \frac{1}{\beta}(1+m\omega_0^2\beta\lambda_0)\partial_q \right] P (q,t)\, ,
\end{equation}
 where the factor in the diffusion coefficient may in this approximation also be written in the form $1/(1-m\omega_0^2\beta\lambda_0)$.

\section{Anharmonic systems}\label{sec:anharmonic}

 Departing from the harmonic oscillator potential, we have to resort to approximation schemes in order to solve the path integral expressions. In the strong friction limit a type of semiclassical or saddle point treatment applies where the coefficients $\lambda_\mu$ serve as expansion parameters. We emphasize that this approach covers the high temperature regime $\gamma\hbar\beta\ll 1$ as well as the challenging deep quantum domain $\gamma\hbar\beta\gg 1$.  In contrast, an alternative formulation developed recently \cite{coffey_2007} is restricted to the former range. For further discussions we refer to Sec.~\ref{sec:comparison}.

 \subsection{Perturbation theory}\label{perturb}

 In the spirit of the previous section we consider the equilibrium density (\ref{eq:impath}) and obtain for its dominating contributions a result of the form
\begin{equation}
\label{eq:semi_dens}
 \rho_\beta (r,x) = \frac{1}{Z} \exp \left[ - \frac{1}{\hbar} S_{E, ma}(r,x) \right] F_{ma}(r,x) \, ,
\end{equation}
with the minimal action $S_{E, ma}$ calculated from the minimal action path $\bar{q}_{ma}(\tau) $. It is given by
\begin{equation}
\label{eq:klq_anharm}
 m \ddot{\bar{q}}_{ma} - V' (\bar{q}_{ma}) - \int_0^{\hbar \beta}d\tau'\,  k ( \tau - \tau' ) \bar{q}_{ma}(\tau') = 0\, .
\end{equation}
subject to the boundary conditions $\bar{q}(0)=r+x/2$ and $\bar{q}(\hbar\beta)=r-x/2$.
Fluctuations around the minimal action path are accounted for by
\begin{equation}
\label{eq:int_fluc}
 F_{ma}(r,x)  =  \int_{y(0)=0}^{y(\hbar \beta) = 0} {\cal D}[y]\,  {\rm e}^{ - \frac{1}{2\hbar} \int_0^{\hbar \beta}d\tau\, y (\tau) \left( \hat{L}  y \right) ( \tau)}
\end{equation}
 containing the second variational order operator
\begin{align} \nonumber
  \hat{L}  y (\tau) = & \left[ - m \partial_\tau^2 +  V'' (\bar{q}_{ma}) \right]  y( \tau) \\
                      &  + \int_0^{\hbar \beta} d\tau'\, k (\tau - \tau')  y (\tau') \, .
\end{align}
In case that there are several minimal action orbits one has to sum over them in (\ref{eq:semi_dens}).
This general expression must now be specified to the situation in the QSR.
The underlying picture there is this: The strong friction suppresses off-diagonal elements of the density matrix meaning that the minimal action path does not deviate much from the mean $ r $ of its initial and final position. Thus, it suffices to  restrict the analysis to the diagonal part $P_\beta(q)=\rho_\beta(q,0)$, i.e.,
\begin{equation}
\label{eq:pbetasemi}
P_\beta(q)=\frac{1}{Z}F_{ma}(q,0)\, {\rm e}^{-S(q)/\hbar}\,
\end{equation}
with  $S(q)=S_{ma}(q,0)$.
The minimal action path and deviations from it are then treated perturbatively in the spirit of a self-consistent approximation.
For this purpose it is convenient to switch to Fourier space
\begin{equation}
\label{eq:q_fourier}
\bar{q}_{ma} (\tau)=\frac{1}{\hbar\beta}\sum_{n=-\infty}^\infty q_n {\rm e}^{i \nu_n \tau}
\end{equation}
leading us from (\ref{eq:klq_anharm}) to
\begin{equation}
\label{fourier_eom}
\nu_n^2 q_n + \gamma  |{\nu_n}|  q_n + v_n /m = b\, .
\end{equation}
Here,
\begin{equation}
\label{eq:fourier_eom}
v_n=\int_0^{\hbar\beta}d\tau \, V'[\bar{q}_{ma} (\tau)] {\rm e}^{-i \nu_n\tau}
\end{equation}
and the term on the right hand side accounts for the fact that the Fourier series periodically continues the path beyond the interval $(0,\hbar\beta)$ producing singularities at times $\tau=n \hbar\beta, n=0, \pm 1, \pm 2, \ldots$ with $b= \dot{\bar{q}}_{ma}(\hbar \beta) - \dot{\bar{q}}_{ma} (0) $.

The strong friction approximation requires systems with sufficiently smooth potentials so that locally the expansion holds
\begin{align}\nonumber
 V'[\bar{q}_{ma} (\tau)] = & V'(q) + V''(q) \left[\bar{q}_{ma} (\tau) - q \right] \\
\label{eq:pot_expansion}
 & + \frac{V'''(q)}{2} \left[ \bar{q}_{ma} (\tau) - q \right]^2 + \dots \, .
\end{align}
 From (\ref{eq:q_fourier}) and (\ref{fourier_eom}) one observes that the $q_n, n\neq 0$ are suppressed at least by factors of order $1/\gamma$, while $q_0$ must be of order 1 due to $\bar{q}_{ma} (0)=\bar{q}_{ma} (\hbar\beta)=q$. Consequently, deviations $\delta q=\bar{q}_{ma} (\tau) - q$ are suppressed by friction as well. Introducing
 an anharmonicity length scale $ l $ the above expansion thus serves as a starting point for a systematic perturbation theory provided that $|\delta q|\ll l$.

 With the above strategy we obtain an explicit expression for the thermal distribution in the overdamped limit. To determine with this result  quantum contributions in the QSE, we follow Sec.~\ref{harm_current} and look for a QSE in the form $\dot{P}=\partial_q\xi  {\cal L}\, P$ with a dynamical factor $\xi$ and a current operator ${\cal L}\sim D_1(q)+\partial_q D_2(q)$. Now, if one writes $P_\beta$ in the form
  \begin{equation}
  \label{eq:pbeta_form}
  P_\beta(q)=\frac{1}{Z}\frac{{\rm e}^{-\psi(q)}}{D_2(q)}\, ,\ \ \psi(q)=\int^q dy \frac{D_1(y)}{D_2(y)}
  \end{equation}
 a vanishing equilibrium current $\left[D_1(q)+\partial D_2(q)\right] P_\beta=0$ is guaranteed in all orders of perturbation theory. Upon comparing the semiclassical expression (\ref{eq:pbetasemi}) with (\ref{eq:pbeta_form}) the most obvious choice is to identify $\psi(q)=S(q)/\hbar$ and $D_2\propto 1/F_{ma}(q,0)$ with a proportionality constant fixed by the high temperature limit. In this range $S(q)\to \hbar\beta V(q)$, while the fluctuation factor tends towards the free particle result $F_{ma}(q,0)\to \sqrt{m/2\pi\hbar^2\beta}$. Hence, with the scaled fluctuation factor
 \begin{equation}
 \label{eq:scaled_f}
 {F}(q)=F_{ma}(q,0)\, \sqrt{\frac{2\pi\hbar^2\beta}{m}}\,
 \end{equation}
 we have the formal expressions for the Moyal coefficients
 \begin{equation}
 \label{eq:moyalfunction}
 D_2(q)=\frac{1}{\beta {F}(q)}\, , \ \ D_1(q)=\frac{1}{\beta {F}(q)}\, \frac{1}{\hbar}\frac{d S(q)}{dq}\, ,
 \end{equation}
 which classically reproduce $D_1(q)\to V'(q)$ and $D_2\to 1/\beta$.

The quantum mechanical current operator is not uniquely fixed by the above choice for the Moyal coefficients though. In fact, any operator of the form
\begin{equation} \label{eq:new_current}
 {\cal L} = d(q) \left[ D_1 (q) + \partial_q D_2 (q) \right]
\end{equation}
obeys ${\cal L}P_\beta=0$ and written as ${\cal L}=\tilde{D}_1+\partial_q \tilde{D}_2$ one can define new Moyal coefficients
\begin{equation}
\label{eq:new_moyal}
 \tilde{D}_1 (q) = d(q) D_1 (q) - d'(q) D_2 (q) \, , \quad \tilde{D}_2 (q) = d(q) D_2 (q) \, .
\end{equation}
Thus, it seems that there is a whole class of Moyal coefficients, where each pair follows from the representative one (\ref{eq:moyalfunction}) by a proper function $d(q)$ obeying $d\to 1$ in the classical limit. By construction all these current operators produce a vanishing current for the  thermal equilibrium (\ref{eq:pbetasemi}). However, based on the previous section the current operator for a harmonic system is known exactly. This information can be used as an additional constraint to fix the correct $d(q)$ as we will show in the sequel. Eventually, dynamical corrections $\xi$ are obtained via an evaluation of the reduced dynamics (\ref{eq:dis_dyn}) within the time window $1/\gamma, \hbar\beta\ll t\ll \gamma/\omega_0^2$.

One remark is in order here: While the representation of the current operator as $\tilde{D}_1+\partial_q \tilde{D}_2$ with the coefficients (\ref{eq:new_moyal})
may be advantageous for  physical interpretations, it has one major drawback compared to (\ref{eq:new_current}) though.
Namely, since $ \tilde{\psi} = \int d y \, \tilde{D}_1 / \tilde{D}_2=S / \hbar -{\rm ln}(d)$ a numerical calculation of $\tilde{\psi}$ based solely on the knowledge of the coefficients $\tilde{D}_{1/2}$ gives an exponent for the equilibrium distribution which carries in addition to the correct action term lower order terms contained in $d$. As $d$ is not known explicitly in this representation the numerically obtained $\tilde{\psi}$ may give rise to spurious equilibrium currents (see e.g.\ \cite{machura_2004}).
Hence, particularly in numerical  applications the representation (\ref{eq:new_current}) is superior.
We will illustrate this by discussing explicit expressions in Sec.~\ref{sec:loc_harm}.

\subsection{Leading order}

In leading order we put in (\ref{eq:pot_expansion}) $V' [\bar{q}_{ma}]=V'(q)$ so that with $v_n\approx V'(q)\hbar\beta \delta_{n,0}$ we find $b^{(0)}=\hbar\beta V'(q)/m$. For $ n \neq 0 $, one has
\begin{equation}
 q_n^{(0)} = \frac{b^{(0)}}{\nu_n^2 + \gamma |\nu_n|}\,
\end{equation}
 and $q_0$ is determined via (\ref{eq:q_fourier}) from $\bar{q}_{ma}(0)=q$. This leads to $q_0^{(0)}/\hbar\beta =q- b^{(0)}\, m\lambda_0\hbar$.
The deviation $\delta q$ can therefore be estimated to be on the order of $(\hbar\beta/\gamma) {\rm ln}(\gamma\hbar\beta)$ for $\gamma\hbar\beta\gg 1$ and on the order of $(\hbar\beta)^2$ for $\gamma\hbar\beta\ll 1$, thus justifying the perturbative treatment.

The corresponding minimal action reads
\begin{equation}
S^{(0)}(q)=\hbar\beta V(q)-\frac{\hbar\beta^2\lambda_0}{2} V'(q)^2\, .
\end{equation}
The fluctuation integral (\ref{eq:int_fluc}) is calculated by replacing $V''[\bar{q}_{ma}]=V''(q)$. The boundary conditions are most conveniently taken into account by introducing $\delta[y(0)]=\delta[\sum_n y_n]$ in the path integral and representing the $\delta$-function as an integral over an auxiliary variable. This way, one gets ${F}^{(0)}(q)=[1-\beta V''(q)\lambda_0]/\beta$ and the thermal distribution in the form $P_\beta^{(0)}=Y^{(0)}\, P_\beta^{(cl)}/Z$ with the quantum correction
\begin{equation}
\label{eq:leadingy}
Y^{(0)}=[ 1-\beta V''(q)\lambda_0]\, {\rm e}^{(\beta^2/2) V'(q)^2 \lambda_0}
\end{equation}
and the {\em unnormalized} classical distribution $P_\beta^{(cl)}=\exp(-\beta V)$ (see also fig.~\ref{fig3} in Sec.~\ref{sec:beyond}).
Now, from (\ref{eq:moyalfunction}) we immediately derive
\begin{equation}
D_1^{(0)}(q)=V'(q)\ , \ \ \ D_2^{(0)}(q)=\frac{1}{\beta}\, \frac{1}{1-\beta V''(q)\lambda_0}\, .
\label{eq:diff0}
\end{equation}
The dynamical factor follows from the path integral calculation as in \cite{ankerhold_2001} and reads as in the classical case $\xi=1/m\gamma$.
 Note that in this order of perturbation theory $D_2^{(0)}(q)\approx [1+\beta V''(q)\lambda_0]/\beta$ so that we regain the QSE already obtained in \cite{ankerhold_2001,ankerhold_qt_semicl}. For a purely harmonic system the result reduces to (\ref{eq:ho_qse_pertur}).

\subsection{Local harmonic approximation} \label{sec:loc_harm}

In next order the full potential is approximated around the endpoints $q$ as a harmonic oscillator meaning that in (\ref{eq:pot_expansion}) all terms beyond the $V''(q)$ contribution are neglected.  From (\ref{fourier_eom}) we obtain
 for $ n \neq 0 $
\begin{equation}
 q_n^{(1)} = \frac{b^{(1)}}{\nu_n^2 + \gamma |{\nu_n}| + \frac{V''(q)}{m}} \, .
\end{equation}
In analogy to the preceeding section, we define a position dependent function
\begin{equation}
 \Lambda(q) = \frac{2}{\beta m} \sum_{n=1}^\infty \frac{1}{\nu_n^2 + \gamma \nu_n + V''(q)/m}  \, ,
 \label{eq:lambda_q}
\end{equation}
which can be expanded according to (\ref{eq:Lambda}) with $\omega_0^2$ replaced by $V''(q)/m$.
Summation over all Fourier components of the minimal action path leads
together with the boundary condition to $ q_0^{(1)} = \hbar \beta q  - \beta b^{(1)} m \Lambda(q)$.
The velocity jump at the endpoint follows as
\begin{equation}
 b^{(1)} = \frac{\hbar \beta \frac{V'(q)}{m} }{1 + \beta V''(q) \Lambda(q)} .
\end{equation}
 Expanding the minimal effective action up to first order in $\delta q=\bar{q}_{ma}(\tau)-q$ yields
\begin{align}
  S^{(1)}(q) = & \frac{m}{2} q b^{(1)}
   + \hbar \beta \left[ V(q) - \frac{q}{2} V' (q) \right]\nonumber\\
 & + \left[ \frac{V'(q)}{2} - \frac{q V''(q)}{2}  \right] \int_0^{\hbar \beta}d\tau \left[\bar{q}_{ma}(\tau) - q \right]\nonumber \\
     = &  \hbar \beta V(q) - \hbar \beta^2  \frac{V'^2(q)}{2} \frac{\Lambda(q)}{1 + \beta V''(q) \Lambda(q)} \, .
     \label{eq:eff_action}
\end{align}
The fluctuation path integral $ F(q) $ is treated accordingly with the second derivative of the potential expanded in the same way as discussed above. In the local oscillator approximation, first and higher order terms in $\bar{q}_{ma} (\tau ) - q $ are truncated consistently. Defining $ u_n = [ \nu_n^2 + \gamma |{\nu_n}| + V''(q)/m ]^{-1} $, one has for the scaled factor
\begin{align} \nonumber
 {F}^{(1)}(q)  = &   \left[ \prod_{n=1}^{+ \infty} \left( \nu_n^2 + \nu_n \gamma \right) u_n \right]
           \left( 1 + 2 \sum_{n = 1}^{+\infty} \frac{u_n}{u_0} \right)^{-1/2} \\
   = & \frac{\exp\left\{-(m \beta/2)\sum_{\mu\geq 1} \lambda_{\mu-1}\, [V''(q)/m]^{\mu}/\mu!\right\}} {\sqrt{1+ \beta V'' (q)  \Lambda(q)}}\, .
   \label{eq:fluc1}
\end{align}
Now, the expressions (\ref{eq:eff_action}) and  (\ref{eq:fluc1}) combine to provide the position distribution in thermal equilibrium for a strongly overdamped system (\ref{eq:pbetasemi}). Since this result is interesting in its own as it applies to all systems with sufficiently smooth anharmonic potentials, it is worth to look for quantum corrections in more detail.
Upon expanding the function $\Lambda(q)$ similarly as in (\ref{eq:Lambda}), quantum mechanical contributions to the classical distribution can be systematically derived. Taking into account terms up to order in $\lambda_0^2$ we have for the quantum correction $Y^{(1)}=Z\, P_\beta^{(1)}/P_\beta^{(cl)}$
\begin{align}
 Y^{(1)}&\approx  \left\{ 1 - \beta  V''(q) \left[\lambda_0 + \frac{3{V''} (q) }{4m} \lambda_1  - \frac{3\beta V''(q) }{4}\lambda_0^2\right]\right\}\nonumber\\
	&\times\, \exp\left\{\frac{\beta^2 V'(q)^2}{2} \left[\lambda_0 +\frac{V''(q)}{m} \lambda_1- \beta V''(q) \lambda_0^2\right]\right\}\, .\nonumber\\
\label{eq:fluc_distribution}
\end{align}
In this expression  $\lambda_1$ and $\lambda_0^2$ dependent terms have been kept so that it applies to the entire strong friction range with $\gamma/\omega_0^2\gg \hbar\beta$. In the high temperature regime $\gamma\hbar\beta\ll 1$ one has $ \lambda_1 \gg m \beta\lambda_0^2 $, while for low temperatures $\gamma\hbar\beta\gg 1 $ the inverse relation holds and $Y^{(1)}$ becomes  a pure perturbation series in powers of $\lambda_0$. In fig.~\ref{fig3} in Sec.~\ref{sec:beyond} the impact of these quantum fluctuations is exemplified.

Before we proceed, we mention that the approach discussed so far can also be extended to include off-diagonal elements of the density matrix. The straightforward calculation leads to
\begin{equation}
\rho_\beta^{(1)}(r,x)=P_\beta^{(1)}(r)\, {\rm e}^{-\Omega(r) x^2/\hbar}
\end{equation}
where
\begin{equation}
  \Omega(q) = \frac{2m}{\beta} \sum_{n=1}^\infty \left[ \gamma \nu_n + \frac{V''(q)}{m} \right] u_n
\end{equation}
corresponds to the local momentum variance. This function can be expanded according to
\begin{equation}
 \Omega(q)  =  \sum_{\mu=1}^\infty \frac{\eta_\mu}{\mu!} \left( \frac{V'' (q) }{m} \right)^\mu\, .
\end{equation}
 For purely ohmic damping $ \eta_0 $ diverges and must be regularized by e.g.\ a Drude cut-off $ \omega_c \gg \gamma $ yielding
\begin{equation}
 \eta_0 \approx \frac{m \hbar \gamma}{\pi} \left[ \Psi \left( \frac{\omega_c}{\nu} \right) - \Psi \left( \frac{\gamma}{\nu} + \frac{\gamma^2}{\nu \omega_c} \right)
	- \frac{\nu}{2 \gamma} + \frac{2 \nu}{\omega_c} \right] \, .
\end{equation}
In the ohmic case the first coefficient stays finite and reads
\begin{equation}
   \eta_1 = \frac{\hbar V''(r)}{\pi \nu} \Psi^{(1)} \left(1 + \frac{\gamma}{\nu} \right)\, .
\end{equation}

Now, that the thermal distribution is at hand, the Moyal coefficients follow according to (\ref{eq:moyalfunction}) with ${F}^{(1)}$ specified in (\ref{eq:fluc1}) and with the action as in (\ref{eq:eff_action}).
To obtain the function $d(q)$ in the ansatz (\ref{eq:new_current}), one observes that for a harmonic system the exponential  in ${F}^{(1)}$ is up to a temperature dependent factor identical with its partition function $Z_0$, while the nominator  is  $\sqrt{\langle q^2\rangle/\langle q^2\rangle_{cl}}$. Thus, the known result (\ref{eq_dia_im}) is regained if
$d^{(1)}=  {F}^{(1)}\, (1+\beta m\omega_0^2 \Lambda) $. The generalization to anharmonic potentials leads then to
\begin{equation}
\label{eq:d_func}
d^{(1)}(q)={F}^{(1)}(q)\, \left[1+\beta V''(q)\Lambda(q)\right]\, .
\end{equation}
Of course, in the classical high temperature domain one has $d^{(1)}=1$.
As discussed around (\ref{eq:new_moyal}), the current operator ${\cal L}=d [D_1+\partial_q D_2]$ with the known $d$ can also be cast into the standard form $\tilde{D}_1+\partial_q \tilde{D}_2$. The result is
\begin{align}
\tilde{D}^{(1)}_1&=\left[1+\beta V''\,\Lambda(q)\right]\frac{d [S^{(1)}-\hbar{\rm ln}F^{(1)}]}{\hbar\beta\, dq}-\frac{d [V''\Lambda(q)]}{dq}\nonumber\\
\tilde{D}^{(1)}_2&=\frac{1}{\beta}\,\left[1+\beta V''\, \Lambda(q)\right]\, .
\label{eq:harm_newmoyal}
\end{align}
With these expressions the current operator in the QSE is completely determined. For practical applications it is more convenient though to systematically expand the above results to the desired order in the inverse friction strength. For this purpose we recall that formally the $\lambda_\mu, \mu\geq 1$ are of order ${\rm ln}(\gamma)/\gamma^{2\mu+1}$, while $\lambda_0$ is of order ${\rm ln}(\gamma)/\gamma$. In particular, in the QSR range $\gamma\hbar\beta\gg 1$ one obtains
\begin{align}
D_1^{(1)}(q)&\approx  V' (q) \left\{ 1+\frac{\beta^2}{4}\lambda_0^2 \left[V''(q)^2+2 V'(q) V'''(q)\right]\right\}\nonumber\\
D_2^{(1)}(q)&\approx  \frac{1}{\beta}\, \frac{1}{1-\beta V''(q)\lambda_0+\frac{3}{4}\beta^2 V''(q)^2 \lambda_0^2}\nonumber\\
d^{(1)}(q)& \approx  1-\frac{\beta^2}{4} V''(q)^2 \lambda_0^2 \label{eq:D_super_1} \, .
\end{align}
Quantum fluctuations appear in the effective drift coefficient
\begin{equation}
\tilde{D}_1^{(1)}(q) \approx V'(q)+\frac{\beta}{2}\lambda_0^2 V'''(q)\left[ V''(q) + \beta V'(q)^2\right]
\end{equation}
 via the anharmonicity of the potential, while in $\tilde{D}^{(1)}_2\approx [1+\beta V'' \lambda_0]/\beta$
only harmonic properties are contained. Corrections here are of order $\lambda_1\ll \lambda_0^2$.

 At this point we come back to our discussion at the end of Sec.~\ref{perturb}. We consider e.g.\ a periodic potential for which one immediately sees that $\psi\equiv S / \hbar$ is periodic as well in each order of perturbation theory for the thermal distribution. For $\tilde{\psi}=\int \tilde{D}_1/\tilde{D}_2$ which must be integrated numerically this is by no means obvious. To consistently neglect in the numerical $\tilde{\psi}$ contributions higher than of order $\lambda_0^2$ is impossible which may thus lead to finite currents in thermal equilibrium. This illustrates why it is advantageous particularly in numerical calculations to work with the coefficients (\ref{eq:D_super_1}) in the current operator.

In principle, the thermodynamic analysis presented above must now be supported by a dynamical calculation of the reduced density (\ref{eq:dis_dyn}) to reveal whether quantum corrections in the dynamical factor $\xi$ are relevant. This can indeed be done within the time window $\hbar\beta, 1/\gamma\ll t\ll \gamma/\omega_0^2$. Effectively, the calculation goes through along the lines described for harmonic systems with $\omega_0^2\to V''(q)/m$. As a result dynamical deviations appear in a form similar as in (\ref{eq:ho_qse}) and are thus completely classical; one has $\xi=1/(m\gamma)+O(V''/\gamma^3)$.  The QSE in this order of perturbation theory is thus obtained as
\begin{equation}
\label{eq:qse1}
\dot{P}(q,t)=\frac{1}{m\gamma}\partial_q {\cal L}^{(1)}\, P(q,t)
\end{equation}
with the current operator defined either in the form (\ref{eq:new_current}) with the  coefficients (\ref{eq:moyalfunction}) and (\ref{eq:d_func}) or in the standard form with (\ref{eq:harm_newmoyal}).

\subsection{Application: Quantum escape rate}

A non-trivial case to prove the consistency of the above QSE is also to calculate the escape rate out of a metastable well and to compare with the semiclassically exact result \cite{wolynes_1981,weiss}.
The latter one can be derived e.g.\ within the Im$F$ approach.
Above the crossover temperature one finds \cite{weiss,ankerhold_qt_semicl}
\begin{equation}
\label{eq:rateimf}
 \Gamma  = \frac{\omega_0}{2\pi}\frac{\omega_R}{\omega_b}\,  f_q {\rm e}^{- \beta V(q_b)}
 \end{equation}
where $\omega_0^2=V''(0)/m$ and $\omega_b^2=|V''(q_b)/m|$ denote frequencies for small oscillations around the minimum of the well at $q=0$ and around the barrier top at $q=q_b$, respectively.
For the quantum enhancement factor one has
 \begin{equation}
 \label{eq:fq_fac}
  f_q  = \prod_{n=1}^{+ \infty} \frac{\nu_n^2+\omega_0^2+\nu_n \gamma}{\nu_n^2-\omega_b^2+\nu_n \gamma}\, ,
\end{equation}
and $\omega_R=-\gamma/2+\sqrt{\omega_b^2+\gamma^2/4}$ is the classical Grote-Hynes frequency. An alternative representation for $f_q$ which is more convenient to compare with strong friction results is given by
\begin{equation} \label{eq:fq}
 f_q = \exp\left[\frac{m \beta}{2}\sum_{\mu=1}^\infty \frac{\lambda_{\mu-1}}{ \mu !}  \left(  \omega_0^{2 \mu} + \left( - 1 \right)^{\mu+1} \omega_b^{2 \mu} \right) \right]\, .
\end{equation}
\begin{figure}
\begin{center}
\includegraphics[width=8.5cm]{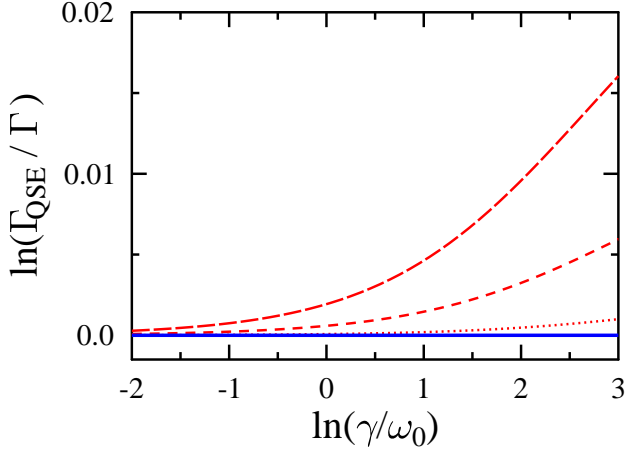}
\end{center}
\vspace*{-0.5cm}
\caption{\label{fig1} (Color online) Escape rates according to the QSE compared to the semiclassical exact rate vs. friction strength for a metastable potential with $\omega_0=\omega_b$ in the high temperature range. The solid line depicts the leading order result [$\lambda_0$-term in (\ref{eq:fq})] for temperatures $\omega_0\hbar\beta=0.2, 0.4, 0.6$; differences between the three data sets are indistinguishable on this scale. Also shown are results obtained according to \cite{coffey_2008} with a friction independent quantum enhancement factor for $\omega_0\hbar\beta=0.2$ (dotted), 0.4 (short-dashed), 0.6 (long-dashed); see text for details.}
\end{figure}
\begin{figure}
\begin{center}
\includegraphics[width=8.5cm]{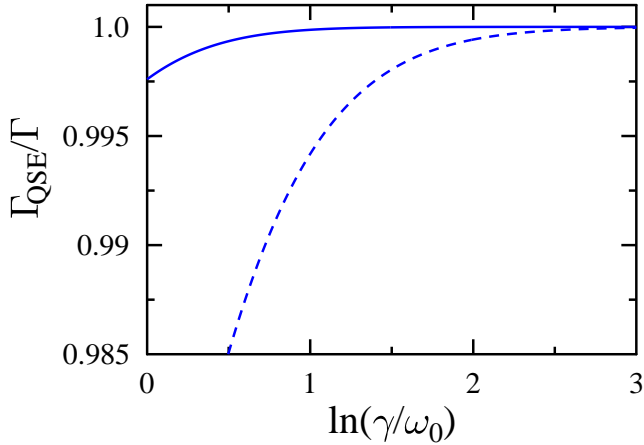}
\end{center}
\vspace*{-2.5cm}
\caption{\label{fig2} (Color online) Same as in fig.~\ref{fig1} but in the low temperature range $\omega_0\hbar\beta=5$ and only for the QSE. The leading order approximation (dashed line) and the next order approximation including terms of order $\lambda_2$ (solid) are shown.}
\end{figure}

The above expression relies on a local harmonic approximation and the anharmonicity of the potential matters only in so far as it leads to a variable curvature when one moves from the barrier top towards the well minimum. This situation thus perfectly fits to the type of perturbative treatment outlined in the previous section with a position dependent diffusion term.
Stationary non-equilibrium solutions of the QSE (\ref{eq:qse1}) follow from ${\cal L}^{(1)} P_\mathit{st} = -J $ with a constant flux $J$. Variation of parameters leads to
\begin{equation}
 P_\mathit{st} (q) = \frac{\gamma m J}{D_2(q)}\, {\rm e}^{-\psi(q)} \int_q^\infty dy \frac{{\rm e}^{\psi(y)}}{d(y)} \, .
\end{equation}
 Thus,  with the normalization $Z=Z_{well}$ of the harmonic oscillator in the well, we gain the escape rate in the form
\begin{align}
 \Gamma_{QSE} = &  J/Z_\mathit{well} \\
        = & \left[ \gamma m \int^{q_b}_{-\infty}dq \frac{{\rm e}^{-\psi(q)}}{D_2(q)} \int_q^{\infty} dy \frac{{\rm e}^{\psi(y)}}{d(y)} \right]^{-1} \, .
\end{align}
Upon evaluating the integrals in saddle point approximation which is justified for sufficiently high barriers, one finds
\begin{equation}
 	\Gamma_{QSE} \approx \frac{\omega_0 \omega_b}{2 \pi \gamma} {\rm e}^{- \beta V(q_b)} f_q \, .
\end{equation}
This is a remarkable result as it reveals that the QSE reproduces the quantum enhancement factor (\ref{eq:fq_fac}) {\em exactly}. The Grote-Hynes frequency corresponds to classical  dynamical corrections and for strong friction reduces to its known leading order approximation $\omega_R\approx 1/\gamma$. Hence, our thermodynamic procedure to derive quantum corrections in the QSE has also been proven {\em a posteriori}.
 In the following subsection we will show that we can extend this method even beyond the local harmonic approximation.

The above findings are illustrated in figs.~\ref{fig1}, \ref{fig2}, where the ratio $\Gamma_{QSE}/\Gamma$ is depicted for different orders of perturbation theory. In the high temperature range (fig.~\ref{fig1}) already the leading order correction of the QSE gives basically the exact result for all damping strength. In contrast, in the approach recently proposed in \cite{coffey_2008} the rate carries the enhancement factor (\ref{eq:fq_fac}) for $\gamma=0$. This leads to increasing deviations for increasing damping strength and/or decreasing temperature and thus reveals the relevance of friction even at elevated temperatures (for further discussions we refer to Sec.~\ref{sec:comparison}).
For low temperatures, see fig.~\ref{fig2}, the QSE gives the correct result for the quantum fluctuations already for moderate friction strengths. Deviations remain small meaning that the perturbative expansion (\ref{eq:fq}) quickly converges. Note that for potentials with $\omega_0=\omega_b$ all odd order contributions cancel.

\subsection{Higher order corrections}\label{sec:beyond}

  We now take into account terms up to second order in the expansion (\ref{eq:pot_expansion}), i.e.\ in the deviations $ \delta q=\bar{q}_{ma}(\tau)-q $ in the equation of motion for the minimal action path (\ref{fourier_eom}). Accordingly,
 one has for the zeroth Fourier component (\ref{eq:fourier_eom})
 \begin{align} \nonumber
 v_0^{(2)}  = & \hbar \beta V'(q) + V''(q) \left( q_0 - \hbar \beta q \right) \\
              & + \frac{V'''(q) }{2} \left[ \hbar \beta \left( \frac{q_0}{\hbar \beta} -q \right)^2
              + 2 \sum_{m= 1}^{ \infty} \frac{ \left| q_m \right|^2}{\hbar \beta} \right] \, ,
\end{align}
while for $n\neq 0$ coupling terms between different modes arise
\begin{align} \nonumber
 v_n^{(2)}  = & \left[ V''(q) q_n + V'''(q) \left( \frac{q_0}{\hbar \beta} -q \right) q_n \right] \\
\label{eq:vn_higher}
        &  + \sum_{m \neq 0, m \neq n} \frac{q_m q_{n-m}}{\hbar \beta} \frac{V'''(q)}{2} \, .
\end{align}
Since these coupling terms can be estimated to be of order
$V'^2 (q) V''' (q)  \lambda_1$, they are negligible against the other contributions for $ \gamma \hbar \beta \gg 1$. Hence,  the leading impact of anharmonicities in the potential can still be treated analytically.
 Note that the term with $|q_m|^2$ in the component $v_0^{(2)}$ is of the same order as the one with $q_m q_{n-m}$ in $v_n^{(2)}$, but appears in (\ref{fourier_eom}) without friction and leads thus to a contribution which is larger roughly by a factor $\gamma$ compared to the former one.
 In complete analogy to our treatment of the minimal action path in the local oscillator approximation, we then find
\begin{align}
 q_n^{(2)} & = \frac{b^{(2)}}{\nu_n^2 + \gamma \left| \nu_n \right| + A(q)/m} \, , \quad n \neq 0 \\ \label{eq:Lambda2}
 \Lambda^{(2)} (q) & = \frac{2}{m \beta} \sum_{n=1}^\infty \frac{1}{ \nu_n^2 + \gamma \left|\nu_n \right| + A(q)/m} \, .
\end{align}
Here, one observes the same structure as in the previous section with the substitution $ V'' (q) \rightarrow A(q) = V''(q) - V'''(q) m b^{(2)} \Lambda^{(2)} / \hbar $.
 Since $\Lambda^{(2)}$ is given by (\ref{eq:Lambda2}) only implicitly, we employ a linearization
\begin{equation}
 \Lambda^{(2)}  = \Lambda(q) + \delta \Lambda \, , \quad  b^{(2)}  = b^{(1)} + \delta b \,
\end{equation}
with $\Lambda(q)$ as defined in (\ref{eq:lambda_q})
to obtain in leading order
 $ \delta \Lambda \approx - \beta V'''(q) V'(q) \lambda_0 \lambda_1 / m$.
Terms of order $\Lambda(q)^3\sim \lambda_0^3 $ and smaller are neglected here, which also implies omission of contributions of order $ \Lambda(q) \delta \Lambda\sim \lambda_0^2\lambda_1\ll \lambda_0^3 $.
In this approximation, the correction of the velocity jump reads
\begin{align} \nonumber
 m \delta b = & - \beta V''(q)  m b^{(1)} \delta \Lambda + \frac{\beta V'''(q)}{2 \hbar} \left[ m b^{(1)} \Lambda(q) \right]^2 \\
              & - \frac{V''' (q) m {b^{(1)}}^2}{2 \hbar} \lambda_1 \left[ 1 + \beta V'' (q) \Lambda(q) \right] \, .
\end{align}
The action $S^{(2)}=S^{(1)}+\delta S$ beyond the local harmonic approximation reads
\begin{align}
 \delta S (q) = & \frac{\hbar \beta^3 {V' (q)}^2 V'''(q)}{2m} \left[ \frac{3}{2} V'(q) - q V'' (q) \right] \lambda_0 \lambda_1 \, ,
 \label{eq:action_2}
\end{align}
which is of the same order of magnitude as $ \delta \Lambda $. Formally, this correction is of order ${\rm ln}^2(\gamma)/(\gamma^4 l) $ with  typical anharmonicity length $l$.

We now turn to the contribution of the fluctuations around the minimal action path $F_{ma}$ and discuss corresponding
anharmonic corrections. In the path integral (\ref{eq:int_fluc}) an
 expansion of the second variational term up to first order in $ \delta q $ leads to an integrand of the form
$ \exp \left[ - \left(\sigma^{(1)} + \delta \sigma \right) \right] $,
where the local harmonic part in the exponent $\sigma^{(1)}$ is given by replacing $ \bar{q}_{ma} (\tau) \to q $ in $V''(\bar{q}_{ma})$ as in the previous section.  The next order term contains the third derivative of the potential
$\delta \sigma  =(V'''/\hbar) \int_0^{\hbar \beta} d \tau \,  y^2 (\tau) [ q(\tau) -q ]$ and reads
\begin{align}  \nonumber
 \delta \sigma  = & \frac{V'''(q) }{\hbar} \left( \frac{q_0}{\hbar \beta} -q \right) \sum_{m = - \infty}^{+ \infty} \frac{|{y_n}|^2}{\hbar \beta} \\
& + \frac{V'''(q)}{\hbar} \sum_{m,n \neq 0} \frac{y_m y_{-m-n}}{\left( \hbar \beta \right)^2} q_n\, .
\label{delta_fluc}
\end{align}
Here, again coupling terms between different Fourier modes appear and lead to a Gaussian path integral which is only tractable numerically. To make progress, we estimate the various terms in (\ref{delta_fluc}) systematically.
For this purpose, we recall that as far as orders of magnitudes are concerned we have $(q_0/\hbar\beta -q)\sim \lambda_0$, $q_n\sim 1/\gamma, n\neq 0$ and $y_n\sim 1/\sqrt{\gamma}, n\neq 0$, $q_0, y_0\sim 1$. The first sum in $\delta\sigma$ produces two types of contributions, namely, $(q_0/\hbar\beta -q) y_0^2\sim \lambda_0$ and $(q_0/\hbar\beta -q) y_{n\neq 0}^2\sim \lambda_0/\gamma\sim {\rm ln}(\gamma)/\gamma^2$. The second sum gives rise to two types of terms as well, one with $y_0 \sum_{n\neq 0} y_n q_{-n}$ and one where the sum contains only $y_m y_{-m-n} q_n$ with $m, n, m+n\neq 0$. This latter part is estimated to be of order $1/\gamma^2$ and thus {\em larger} than the action correction in (\ref{eq:action_2}). The former one generates a shift in the Gaussian integral over $y_0$ which after performing the integration leads to a contribution with $(\sum_{n\neq 0} y_n q_{-n})^2\sim 1/\gamma^3$. The essence of this analysis is that if only the first sum in $\delta\sigma$ is kept, the distribution is calculated with corrections at most of order ${\rm ln}(\gamma)/\gamma^2$. This in turn means that the correction $\delta S$ in the action must be neglected.
This way, one again ends up with independent Gaussian integrals over the fluctuations modes $y_n$, the local frequency  $A(q)$ of which reads in this order of perturbation theory
\begin{align}
A&\approx  V''-\beta V' V'''\left[\lambda_0  -\beta V''\lambda_0^2 \right]\, .
\label{eq:ampli_a}
\end{align}
The result for the scaled fluctuation factor (\ref{eq:scaled_f})  $F^{(2)}$ follows thus from $F^{(1)}$ in (\ref{eq:fluc1}) with the replacement $V''(q)\to A(q)$. Contributions to the full path integral beyond the Gaussian approximation for the $y_n$ are also negligible.  Hence, the quantum contribution to the thermal distribution $Y^{(2)}=Z\, P^{(2)}_\beta/P^{(cl)}_\beta$ in the range $\gamma\hbar\beta\gg 1$ reads
\begin{align} \label{eq:Y2}
 Y^{(2)} &\approx  \left\{ 1 - \beta  V'' \lambda_0 + \beta^2 \left[ \frac{3}{4} {V''}^2 + V' V'''  \right]\lambda_0^2\right\}\nonumber\\
 &\times {\rm e}^{-[S^{(1)}-\hbar\beta V]/\hbar}
\end{align}
where the action contribution $S^{(1)}-\beta\hbar V$ is the exponent in (\ref{eq:fluc_distribution}).
This result is exact up to order $ \log(\gamma)^2 / (\gamma^2 l) $.
Higher order correction terms couple the Fourier modes in the minimal action path as well as in the fluctuation path integral and are analytically no longer accessible.

The above distribution determines the Moyal coefficients of the
QSE to be of the form
\begin{align} \nonumber
D_1^{(2)} \approx & V'(q) \left\{ 1 + \frac{\beta^2}{4} \left[ {V'' (q) }^2 - 2 V'(q) V'''(q) \right] \lambda_0^2 \right\} \\
\label{eq:moyal_2}
  D_2^{(2)} \approx & \frac{1}{\beta} \frac{1}{ 1 - \beta  V''(q) \lambda_0 + \beta^2 \left[ \frac{3}{4} V''(q)^2 + V'(q) V'''(q)  \right]\lambda_0^2} \, .
\end{align}
Now, also the diffusion coefficient carries information about the anharmonicity [cf.~(\ref{eq:D_super_1})]
Interestingly, the structure of the drift $D_1^{(2)}$ is similar to that of $D_1^{(1)}$ with $2 V' V'''$ replaced by $-2 V' V'''$. This is a direct consequence of the perturbation theory for the position distribution, which reproduces $P_\beta^{(k)}$ only with the given set of $D_1^{(k)}, D_2^{(k)}$. Apparently, the leading anharmonic corrections to $P_\beta$ appear in the fluctuation prefactor only.
Eventually, we replace in $ d (q) $ the second derivative of the potential consistently by $ A (q) $ so that
\begin{equation}
 d^{(2)} (q) \approx 1 - \frac{\beta^2}{4} {V'' (q) }^2 \lambda_0^2
\end{equation}
remains unaltered compared to $d^{(1)}$. Upon inspection of (\ref{eq:Y2}) and (\ref{eq:moyal_2}), we find the third derivative of the potential to appear in the second order term in $ \lambda_0 $. This reflects the
nonlocal feature of the deep quantum regime and shows that $ \lambda_0 $ indeed plays the role of a semiclassical parameter similar to Planck's constant for an isolated quantum
system. In principle, one must now also extract dynamical corrections from the reduced dynamics. However, we know from the previous section that in local harmonic approximation these are at most of order $1/\gamma^2$ compared to the leading term $1/m\gamma$. The quantum corrections accounted for in the above coefficients are much larger, namely, of order ${\rm log}(\gamma)^2/(\gamma^2 l)$. As a consequence, additional dynamical correction due to the anharmonicity do not play a role.

\begin{figure}
\vspace*{0.5cm}
\begin{center}
\includegraphics[width=8.5cm]{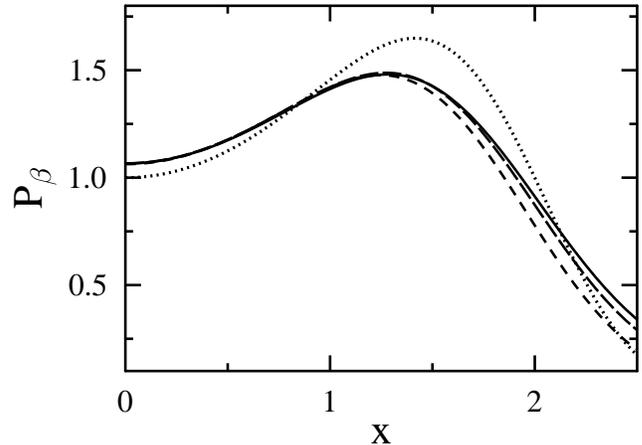}
\end{center}
\vspace*{-0.5cm}
\caption{\label{fig3} Equilibrium distribution scaled with $P_\beta^{(cl)}(x=0)$ in a double well potential vs.\ the scaled position $x=q/\sqrt{\hbar/m\omega_0}$ for various orders of perturbation theory. The classical result (dotted) is depicted together with the leading order expression (short-dashed) according to (\ref{eq:leadingy}), the expression in local harmonic approximation (dashed) according to (\ref{eq:fluc_distribution}), and the expression beyond (solid) according to (\ref{eq:Y2}). Parameters are $\gamma/\omega_0=3$, $\omega_0\hbar\beta=1$ for the bath and $\alpha\hbar/m^2\omega_0^3=0.5$ for the potential so that the well minimum is located at $x= \sqrt{2}$.}
\end{figure}

To illustrate the perturbative results for the equilibrium distribution, we consider a double well potential of the form $V(q)=-m \omega_0^2\, q^2/2+\alpha\, q^4/4$. In Fig.~\ref{fig3} the thermal density is shown in the classical regime together with the quantum distributions in leading order, local harmonic approximation, and beyond  as discussed above.
The leading order describes the quantum tunneling in the barrier range, while the higher order approximations also capture the tunneling at the rising walls of the potential. Around the barrier top all quantum results basically coincide and start to deviate only towards the well region. Further, the maximum of the distribution shifts slightly towards the barrier top thus reflecting the finite transparency of the barrier.

\section{Comparison with other approaches}\label{sec:comparison}

The analysis of the strong friction limit in quantum mechanics based on path integrals as outlined in \cite{ankerhold_2001}  has triggered other studies  proposing alternative approaches to derive quantum Smoluchowski equations.
Here, we briefly discuss them in comparison with this previous work and its extension presented above. For this purpose we recall that one carefully has to distinguish between the strong friction range at high temperatures $\gamma\hbar\beta\ll 1$ and that at low temperatures $\gamma\hbar\beta\gg 1$. While in the latter domain the strong friction analysis {\em must}  be based on a formulation which captures the system-bath interaction non-perturbatively, in the former region one could hope that perturbative formulations like e.g.\ master equations are sufficient to derive at least leading quantum corrections. This assumption is based on the fact that the Wigner representation of e.g.\ the Caldeira-Leggett master equation reduces in the high temperature limit to the classical Fokker-Planck equation.

One approach put forward in \cite{bolivar} starts directly from the classical Smoluchowski equation and quantizes this reduced equation of motion. Such a procedure is questionable already at sufficiently elevated temperatures and  it is certainly not a consistent way to take into account the quantum mechanics of the reservoir and its interaction with the system. In fact, corresponding results contradict the fluctuation-dissipation theorem and may thus even lead to unphysical predictions. This failure has already been discussed in \cite{bolivar_comment}.

The formulation developed in \cite{tsekov_2007} focuses on the case of a QSE for free Brownian motion and uses an {\em ad hoc} procedure to derive quantum corrections in the mean square displacement in position.
From the exact path integral results \cite{weiss,grabert_qbm} one finds that  in this case $\langle [q(t)-q(0)]^2\rangle= 2 (t +\Delta t)/M\beta\gamma$, where $\Delta t$ captures deviations from the classical behavior.  For strong friction and on the coarse grained time scale $\hbar\beta, 1/\gamma\ll t$ [see (\ref{eq:smol_reg})]  $\Delta t$ is at most of order $\hbar\beta$ and  must thus be discarded in a systematic treatment. As shown above, the QSE for a free Brownian particle coincides with the classical Smoluchowski equation.

In the high temperature domain $\gamma\hbar\beta\ll 1$ Coffey and co-workers \cite{coffey_2007,coffey_2008} have recently derived quantum corrections in a QSE by determining a current operator of the form (\ref{eq_dia_im}), however, using the  thermal Wigner function of the {\em bare} system.  The crucial question is, whether in this domain quantum corrections can then be derived systematically. Unfortunately, and in contrast to a naive expectation, this is not the case. To see this in detail, we look at the exactly solvable harmonic system treated already in Sec.~\ref{sec:harmonic}. In the strong friction limit the exact diffusion term in the QSE is given by the variance in position $m\omega_0^2 \langle q^2\rangle$ as shown in (\ref{eq:ho_qse}), i.e.,
\begin{equation}
\label{q2add}
D_2^{\rm ex}  =  \frac{\omega_0^2}{\beta} \sum_{n=- \infty}^{+ \infty} \frac{1}{\nu_n^2 +  |\nu_n| \gamma + \omega_0^2}\, .
\end{equation}
 For $\gamma\hbar\beta\ll 1$ this expression can be expanded in powers of $\hbar$. In leading order, only the $n=0$ contribution must be taken into account and
one regains the classical result. In next order, only the $\nu_n^2, n\neq 0$ term in the denominator must be kept due to $\gamma\hbar\beta\ll 1$, which in turn implies $\omega_0\hbar\beta\ll 1$ since $\gamma/\omega_0\gg 1$. Hence, the leading quantum correction is $(\omega_0^2/\beta) \sum_{n\neq 0} (1/\nu_n^2)=\omega_0^2\hbar^2\beta/12 $. This is the well-known universal quantum correction, also discussed in textbooks \cite{weiss}, which reveals that in the high temperature limit the leading quantum correction is independent of friction. Hence, in principle it can indeed also be derived from the {\em bare} equilibrium distribution.

Now let us look at higher order quantum corrections. An $\hbar$-expansion of the Wigner function of the {\em bare} system as done in \cite{coffey_2007,coffey_2008} leads to a powers series exactly with  $(\beta/2)\times\hbar^2\beta/12 $ as expansion parameter. Accordingly, quantum corrections in a diffusion coefficient gained with this expansion are independent of $\gamma$ to all orders of $\hbar$. This can only be true though if the $\nu_n\gamma$ term in the above sum can be neglected in {\em all} orders in $\hbar$. This leads to the conditions $ \nu_n\gamma\ll \nu_n^2, \omega_0^2\, , n\neq 0 $. The first relation brings us back to $\gamma\hbar\beta\ll 1$, while the second one gives $\gamma/\omega_0\ll \omega_0\hbar\beta$. For strong friction $\gamma/\omega_0\gg 1$ these two conditions {\em contradict} each other since the first one implies
 $\omega_0\hbar\beta\ll 1$, while the second one requires $1\ll \omega_0\hbar\beta$. It is thus {\em not} consistent to neglect the $\gamma$-dependence of the diffusion term beyond the leading quantum correction.
 The correct expansion in the strong friction range which only requires $\gamma/\omega_0^2\gg \hbar\beta$ to ensure the validity of the Markov approximation and applies from $\gamma\hbar\beta\ll 1$ to $\gamma\hbar\beta\gg 1$ is given in (\ref{eq:Lambda}). In the high temperature range $\omega_0\hbar\beta\ll\gamma\hbar\beta\ll 1$ this leads to the series
 \begin{equation}
 D_2^{\rm ex}=\frac{1}{\beta}\left[1+\frac{(\omega_0\hbar\beta)^2}{12}-\frac{\gamma}{\omega_0}\frac{(\omega_0\hbar\beta)^3}{4 \pi^3}\zeta(3)+\ldots\right]
 \end{equation}
 in contrast to Coffey's result
 \begin{equation}
 D_2^{\rm Coff}=\frac{1}{\beta}\left[1+\frac{(\omega_0\hbar\beta)^2}{12}-\frac{(\omega_0\hbar\beta)^4}{720}+\ldots\right]\, .
 \end{equation}
 The first conclusion is that Coffey's approach is correct in the regime $\gamma\hbar\beta\ll 1$ {\em only} in leading order in $\hbar$, where it coincides with the leading order result presented above. Higher order corrections in the QSE as specified e.g.\ in \cite{coffey_2008} are incorrect and, while they may give qualitatively reasonable results, lead to uncontrolled approximations. The second and more fundamental conclusion is that for larger friction sub-leading quantum fluctuations are determined by the system-bath interaction so that any approach which treats this interaction perturbatively fails even at elevated temperatures. 
 
 In the above discussion we focused on the position variance for a purely ohmic spectral density. In a strict sense, however, the momentum variance diverges in this limit [cf. discussion below (\ref{momentum})]. This problem is usually cured by introducing a high frequency cut-off $\omega_c$ as mentioned at the end of Sec.~\ref{dissdyn} \cite{weiss}. The above high temperature expansion remains then valid apart from corrections of order $\gamma/\omega_c, \omega_0/\omega_c$ if $\omega_c\hbar\beta\gg 1$. In particular, this latter relation
 guarantees that equilibrium fluctuations are still determined by almost ohmic spectral densities and almost ohmic friction functions $\hat{\gamma}(z)$ in accordance with an effectively Markovian theory.

\begin{figure}
\begin{center}
\includegraphics[width=8.5cm]{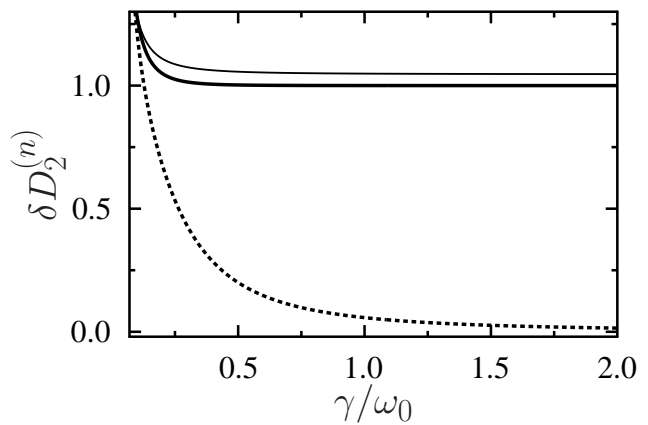}
\end{center}
\vspace*{-0.5cm}
\caption{Quantum corrections $\delta D_2^{(n)}$ (\ref{fidel}) in the diffusion coefficient for a harmonic system for $n=1$ (thin) and $n=2$ (thick).  The systematic strong friction expansion according to (\ref{eq:Lambda}) (solid) is shown together with the ''without-$\gamma$-expansion'' \cite{coffey_2008} (dotted). The parameter $ \gamma \hbar \beta = 0.4 $ is kept fixed, while $ \gamma / \omega_0 $ is varied. For  $n=1$ both expansions coincide (see text).}\label{fig:fidelities}
\end{figure}
To illustrate this discussion we show in fig.~\ref{fig:fidelities} successive quantum corrections in the exact diffusion coefficient $D_2^{(\rm ex)}$  of a harmonic system according to (\ref{eq:Lambda}) and according to Coffey's ''without-$\gamma$-expansion'', respectively. It is convenient to consider the respective $n$th order approximation $D_2^{(n)}$ properly weighted by the full result, i.e.,
\begin{equation}
\label{fidel}
 \delta D_2^{(n)} = \frac{D_2^{(n)} - D_2^{(n-1)}}{D_2^{\rm ex} - D_2^{(n-1)}} \, .
\end{equation}
For $n=1$ we have  $D_2^{(0)}=D_2^{(\rm cl)}$ and $D_2^{(1)}=D_2^{(\rm cl)}+\omega_0^2 \hbar^2\beta/12 $ so that both expansions coincide as pointed out above. Substantial differences appear, however, for $n=2$ in the range, where the authors of \cite{coffey_2007} claim their approach to work, namely, for $\gamma\hbar\beta\ll 1$ and $\gamma/\omega_0\gg 1$. Even in the strongly underdamped regime for fixed $\gamma\hbar\beta\ll 1$ the ''without-$\gamma$-expansion'' is only of limited use since one then leaves the classical regime and enters the low temperature range.

The escape rate in the high temperature range obtained within this type of approximation is shown in fig.~\ref{fig1}. According to the above discussion the corresponding quantum enhancement factor is found to be independent of friction \cite{coffey_2008}. Obviously, deviations tend to increase for stronger friction, i.e.\ in the regime where the approach is supposed to apply.

\section{Discussion}\label{sec:discussion}

In this paper we developed a systematic semiclassical expansion for the imaginary time path integral of the statistical operator in order to derive drift and diffusion terms in the QSE. Since in the QSR the thermal time scale $\hbar\beta$ is still small compared to the time scale for relaxation of the marginal distribution in position, non-Markovian features of the reduced dynamics do not play a role and a QSE exists. Drift and diffusion are directly related to the action of the extremal paths and fluctuations around them. The physical picture behind is that for strong friction  and on the thermal time scale $\hbar\beta$ orbits explore the potential only locally. This allows for a perturbative treatment which successively takes higher order terms in the expansion of the potential around the average minimal action orbit into account.
Hence, in leading order the dynamics takes effectively place  in a potential of constant force, in next order in a local harmonic potential, and anharmonicities become relevant only beyond. Dynamical corrections are classical in nature, at least up to the local harmonic approximation.

The systematically improved QSE is applied to calculate the escape rate out of a metastable state for low temperatures, but above crossover. It then reproduces the quantum enhancement factor exactly, whereas dynamical corrections only appear in the classical form of the Grote-Hynes frequency. Note that for strong friction the inverse crossover temperature $\beta_0$ is given by $\hbar\beta_0\approx 2\pi \gamma/\omega_b^2$. On the other hand the QSR condition requires $\hbar\beta\ll \gamma/\omega_b^2$. This is why the deep tunneling regime below the crossover lies outside the range of validity of the QSE and can thus not be captured by this type of approximation.
Above the crossover, however, the rate expression is not sensitive to local anharmonicities of the metastable potential,
since such contributions only enter in third  and higher
order terms of a variational expansion around the trivial minimal action paths residing at the well bottom and at the barrier top.

The systematic expansion outlined here allows to study quantum effects in overdamped systems in a broad range of parameters comprising the domains of {\em high} as well as of {\em low} temperatures and due to the high order of perturbation theory also the range of strong to only moderate dissipation.

\section*{Acknowledgments}
We thank H. Grabert and A. Verso for fruitful discussions.

\bibliography{biblio}

\end{document}